\documentclass[12pt,preprint]{aastex}

\usepackage{graphicx}

\def\pn{\par\noindent}

\def\cgs{erg cm$^{-2}$ s$^{-1}$} 
\def\lum{erg s$^{-1}$} 
\def\cm{cm$^{-2}$} 
\def\gsimeq{\hbox{\raise0.5ex\hbox{$>\lower1.06ex\hbox{$\kern-1.07em{\sim}$}$}}} 
\def\lsimeq{\hbox{\raise0.5ex\hbox{$<\lower1.06ex\hbox{$\kern-1.07em{\sim}$}$}}}

\slugcomment{Submitted to ApJ}

\shorttitle{Suzaku observations of hard X--ray selected Seyfert galaxies}
\shortauthors{Comastri et al.}

\begin{document}


\title{Suzaku observations of hard X-ray selected Seyfert 2 galaxies}


\author{A. Comastri\altaffilmark{}, K. Iwasawa\altaffilmark{}, R. Gilli\altaffilmark{}}
\affil{INAF-Osservatorio Astronomico di Bologna, 
    via Ranzani 1, 40127 Bologna ITALY}
\email{andrea.comastri@oabo.inaf.it}

\author{C. Vignali\altaffilmark{}, P. Ranalli\altaffilmark{} }
\affil{Dipartimento di Astronomia Universit\`a di Bologna, 
via Ranzani 1, 40127 Bologna ITALY} 

\author{G. Matt\altaffilmark{}}
\affil{Dipartimento di Fisica, Universit\`a di Roma Tre,
via della Vasca Navale 84, 00146 Roma ITALY}

\author{F. Fiore\altaffilmark{}}
\affil{INAF-Osservatorio Astronomico di Roma,
 via Frascati 33, 00040 Monteporzio Catone, ITALY}

\begin{abstract}

We present {\it Suzaku} observations of five hard X--ray selected  nearby Seyfert 2 galaxies.
All the sources were clearly detected with the {\tt pin} Hard X--ray Detector 
up to several tens of keV, allowing for a fairly good characterization of the broad--band 
X--ray continuum. We find that a unique model, even including multiple components, 
fails to represent the spectra of all the sources. Heavy obscuration manifests itself 
in different flavours. For two sources there is evidence for a reflection dominated 
continuum; among the other three, one is ``mildly" Compton thick ($N_H \sim$ 10$^{24}$ \cm), 
while the remaining two are heavily obscured ($N_H \simeq 10^{23.5}$ cm$^{-2}$), 
but Compton thin. 
Strong, narrow, iron K$\alpha$ lines ($EW \sim$ 1--2 keV) due to neutral or mildly ionized gas,
are detected in Compton thick AGN.
In all of them the K$\alpha$ line is accompanied by the K$\beta$. 
The intensity and shape of the soft X--ray spectrum are different from 
object to object. Soft X--rays may  originate from a nuclear component 
scattered off, or leaking through, the X--ray absorber, plus 
thermal X--rays from the host galaxy. 
Emission from circumnuclear gas photoionized by the active nucleus, parameterized 
with a power law plus individual narrow Gaussian lines, also provides 
an acceptable description of the soft X--ray spectra.
The limited {\it Suzaku} {\tt XIS CCD} energy resolution does not allow us to draw firm 
conclusions on the origin of the soft X--ray emission. 
We briefly discuss our findings in the light of AGN Unified model and 
the geometry of the obscuring gas.

\end{abstract}


\keywords{galaxies: active --- X--rays: observations --- Seyfert galaxies:
individual (NGC 5728, NGC 4992, ESO263--G13, ESO 137--G34, ESO323--G32) }



\section{Introduction}

A fraction as high as 50\% of Seyfert 2 galaxies in the nearby Universe are
obscured in the X--ray band by column densities of the order of, 
or larger than the inverse of the Thomson cross-section 
($N_H\ge \sigma_T^{-1} \simeq 1.5 \times 10^{24}$~cm$^{-2}$),  
hence dubbed Compton thick (CT). If the optical depth ($\tau = N_H \sigma_T$) 
for Compton scattering does not exceed values of the order
of ``{\it a few}", X--ray photons with energies higher than 10--15 keV are able 
to penetrate  the obscuring material and reach the observer. For higher values
of $\tau$, the entire X--ray spectrum is depressed by Compton down scattering 
and the X--ray photons are effectively trapped by the obscuring material 
irrespective of their energy. The former class of sources (mildly CT) can be 
efficiently detected by X--ray instruments sensitive above 10 keV, while for the 
latter (heavily CT) their nature may be inferred  through indirect arguments, 
such as the presence of a strong iron K$\alpha$ line over a flat reflected continuum. 
The search for and the characterization of the physical 
properties of CT AGN is relevant to understand the 
evolution of accreting Supermassive Black Holes (SMBHs). In particular,  
mildly CT AGN are the most promising candidates to explain the 
so far largely unresolved spectrum of the X--ray background 
around its 30 keV peak (Worsley et al. 2005; Treister \& Urry 2005; Gilli et al. 2007). 
According to the Gilli et al. (2007) XRB synthesis model, their integral
contribution to the hard X--ray background is of the order of 25--30\%.
This fraction has been estimated under simplified hypotheses. 
In particular, the same luminosity function and cosmological evolution 
of unobscured and Compton thin AGN is assumed; moreover, the number
density of ``heavily" CT AGN is the same of ``mildly" CT. 
While these assumptions are not inconsistent with the 
present observational framework (Risaliti et al. 1999; Guainazzi et al. 2005),
it should be noted that absorption column densities  in excess of 10$^{24}$ \cm\ 
were measured or inferred for about a few tens of nearby AGN 
(Comastri 2004; Della Ceca et al. 2008), and only a handful of 
them are known beyond the local Universe (Norman et al. 2002; Iwasawa et al. 2005).
CT AGN may represent a significant fraction of the 
accretion power in the Universe and indeed a correction for 
the CT contribution to the estimates of the 
SMBH local mass density has to be included in the 
calculations (Marconi et al. 2004).

An unbiased census of extremely obscured AGN would require to survey the hard 
X--ray sky above 10 keV with good sensitivity. Such an argument is one of the
key scientific drivers of the {\sc nustar} (Harrison et al. 2010) and   
{\sc astro-h} (Takahashi et al. 2010) missions, which will be launched in the next few years,
and of the {\sc nhxm} mission study (Pareschi et al. 2010). 

For the time being, one has to rely on the observations obtained by 
the high--energy detectors on board {\sc bepposax}, {\sc integral}, {\sc swift}
and, more recently, {\it Suzaku}. All--sky surveys were performed 
using both the IBIS coded-mask telescope onboard {\sc integral} 
and the BAT detector on board {\sc swift}.
Though limited to bright and thus low--redshift sources, they have proven to be quite 
successful. More than one hundred AGN are reported in both {\sc integral}
(Beckmann et al. 2009) and {\sc swift}  catalogues (Tueller et al. 2010; Cusumano et al 2010).

Hard X--ray selection is less biased against absorption and thus 
provides a useful benchmark to study the column density distribution 
of obscured AGN and the fraction of CT sources 
in the local Universe. 
The large majority of {\sc Integral} and {\sc Swift} sources were observed
by XMM and {\it Chandra} and their  broad band X--ray spectra 
discussed by Winter et al. (2008). 
While the already known, nearby  CT AGN were recovered 
by {\sc Swift} and {\sc Integral} surveys, the fraction of ``newly" discovered 
CT AGN is surprisingly low and apparently inconsistent, by about a factor of 2, 
with that predicted by Gilli et al. (2007) in the local Universe.
On the basis of these findings, it has been 
proposed (Treister et al. 2009) that the contribution of 
CT AGN to the hard X--ray background may be significantly lower 
(by a factor of 2 to 3) than previously thought, with important implications 
for the evolution of the accretion power.

Since absorption column densities are often measured on relatively poor 
quality X--ray spectra taken from non-simultaneous observations, above 
and below about 10 keV, and combining 
different instruments, the CT fraction and, more in general, the absorption 
distribution in the local Universe may still be subject to several uncertainties. 
Good quality simultaneous spectra extending over the 0.5--100 keV energy range are needed
for a robust measurement of absorption column densities, especially in the 
Compton thick regime. Moreover, nearby obscured AGN always show excess 
emission above the extrapolation of the obscured nuclear spectrum (e.g. Turner et al. 1997).  
Soft X--ray spectroscopy is a powerful tool to study the origin of this component 
which is plausibly related to warm gas photoionized by the nuclear 
continuum (Guainazzi \& Bianchi 2007).  
The X--ray detectors onboard the Japanese satellite 
{\it Suzaku} are well suited to this purpose, at least as far as ``local" AGN are 
concerned.
We have conceived a program with {\it Suzaku} to observe nearby, 
relatively X--ray bright ($> 10^{-11}$ \cgs) AGN selected by requiring {\bf i)} a significant 
detection above 10 keV by either {\sc Swift} or {\sc Integral} and {\bf ii)} evidence for 
significant X--ray absorption at lower energies from archival observations.
The immediate science objective 
is the characterization of the broad band X--ray spectra of 
hard X--ray selected, heavily obscured AGN and a first step 
towards a better census of Compton thick 
absorption in the nearby Universe.

\section{Sample selection}

Five sources (NGC 4992, NGC 5728, ESO137--G34, ESO263--G13, ESO323--G37) 
spectroscopically classified as Seyfert 1.9--2.0 galaxies,
were originally selected from the {\sc integral/ibis} 
(Beckmann et al. 2006) and {\sc swift/bat} (Markwardt et al. 2005) catalogues. 
The column densities, as inferred from archival ({\it Chandra} and XMM--{\it Newton})
observations, are of the order of $10^{23-24}$ cm$^{-2}$, though affected by large errors. 
By construction, the sample is biased towards obscured AGN.

The extremely bright flux of NGC 4992 in the {\sc integral} observation, coupled with the
lack of a detection in the RASS (Voges et al. 1999), is consistent with an absorbed spectrum, 
as confirmed by a snapshot {\it Chandra} observation (Sazonov et al. 2005).
\par
{\it Chandra} revealed an heavily obscured nucleus in NGC 5728 (Zheng et al. 2006). 
Although the column 
density is not well constrained due to the limited response at high energies, the presence
of a strong iron line (EW $\sim$ 1 keV) suggests that the source might be obscured
by Compton thick gas.
\par 
ESO137--G34 is detected by {\sc Integral} and subsequently observed 
with XMM-{\it Newton} (Malizia et al. 2009). The combined XMM and {\sc Integral} spectrum 
can be well fitted by both a transmission and a reflection model as well as by a more 
complex 
absorption distribution. In all cases, there is evidence for Compton thick absorption. 
\par
ESO323--G32 and ESO263--G13 were selected on the basis of a significant
($>5 \sigma$) {\sc integral/ibis} detection, unambiguous optical classification 
as Seyfert 2 galaxies, and the lack of archival observations in the 2--10 keV 
energy range. Both are detected by {\sc rosat} with a soft X--ray spectrum.
The extrapolation of the best--fit {\sc integral} spectrum
in the soft {\sc rosat} band suggests obscuration 
in excess of 10$^{23}$ cm$^{-2}$.

\section{Observations and data Analysis}

The fifth Japanese X--ray satellite {\it Suzaku} (Mitsuda et al. 2007) carries four 
X--ray telescopes with a CCD camera, the X--ray imaging spectrometer ({\sc XIS}; Koyama et al. 2007) 
in the focal plane of each of them, and a non--imaging Hard X--ray Detector 
({\sc HXD}; Takahashi et al. 2007), consisting of Silicon
{\tt pin}  photo-diodes and {\sc GSO} scintillation counters. They simultaneously cover 
the 0.2--12 keV and 10--700 keV energy ranges, respectively, although 
with different sensitivities. 

The XIS cameras were operating with the Space-Row Charge Injection in 
the standard mode. The data were taken by the three XISs (two FI CCD 
cameras -- XIS0 and XIS3 --  and one BI CCD camera XIS1) and the HXD pin. 
For the two sources -- NGC 5728 and NGC 4992 -- observed 
at the beginning of the mission, data from the FI CCD XIS2 are also included. 
All the sources were observed in the 
``HXD" nominal position on the detector plane which is optimised for the 
maximum throughput of the HXD rather than the XIS. 

The data from version 2 processing\footnote{http:// www.astro.isas.ac.jp/suzaku/process/caveats}
 were used for the analysis. The cleaned event files produced by the data processing with 
the standard selection criteria were used. The non X-ray background of the 
{\tt pin} detector was estimated by the ``tuned" background model. 
The effective exposure times and the background subtracted count rates 
in the {\sc xis} and {\sc pin} detectors for each object are reported in Table 1.

Spectral fits are performed with XSPEC version 11.3.
Front Illuminated CCD were co-added, while the back illuminated chip was discarded
after verifying that its spectrum does not significantly improve the quality of 
spectral fits. 

The data from XIS and {\tt pin} detectors were fitted together. 
The intercalibration constant ({\tt pin}/{\tt XIS}) is 
1.18\footnote{ftp:// legacy.gsfc.nasa.gov/suzaku/doc/xrt/ suzakumemo-2008-06.pdf}.
The broad band {\it Suzaku} spectra in a $\nu$F$_{\nu}$ representation 
of the five sources are shown in Figure~1. 
In Figure 2 we report the XIS spectra.

\section{Spectral analysis}

For sake of clarity and to ease the comparison with some of the previous {\it Suzaku} results 
on heavily obscured AGN (Ueda et al. 2007; Eguchi et al. 2009), we consider two baseline 
spectral models for the broad band {\it Suzaku} spectra, namely  
a transmission dominated (TD) model: {\tt wa$_{Gal}$(apec + pexrav + zpcafbs(cutoffpl+2$\times$zgauss))} 
in {\sc XSPEC}
notation   and a reflection dominated (RD) model: {\tt   wa$_{Gal}$(apec + pow + pexrav + 
2$\times$zgauss)}. 
The spectral components of the TD model are: a power law with an exponential cut-off 
fixed at 200 keV (Gilli et al. 2007) for the primary spectrum plus intrinsic 
absorption, a reflection component from neutral gas ({\tt pexrav}; Magdziarz 
\& Zdziarski 1995) and iron K emission (both K$\alpha$ and K$\beta$, whenever 
required by the spectral fits). 
The photon index of the reflection component is linked to that of the primary continuum and 
no absorption on the reflected spectrum is considered.
The RD model includes the same spectral components except for the 
absorbed primary power law. The photon index of the RD model is left free to vary.
The intensity of the reflection component $R$ in TD model 
is defined as the ratio between the normalization of the reflected and primary 
(i.e. power law) spectrum. The inclination angle of the reflection component 
is fixed to $cosi$=0.9, which is 
equivalent to assume a rather face--on geometry for the reflecting material, 
likely to be associated with the inner walls of the obscuring torus.

Soft X--ray emission in excess of the extrapolation of
the absorbed nuclear continuum at low energy is observed in all the 
sources of our sample. This emission may be due to hot gas associated 
to the host galaxy and is modeled with an updated 
version of the original Raymond \& Smith (1977) code for thermal plasmas 
({\tt apec} in {\tt Xspec}), and to AGN primary continuum  scattered 
on the line of sight or leaking through the nuclear absorber, represented 
by a partial covering ({\tt zpcfabs in XSPEC}), in the TD model, and with a power law in the RD model.
In both models the power law slope is assumed to be the same of the intrinsic hard X--ray 
continuum, with the exception of one object (see $\S$ 4.4). 
The scattering fraction $f$ can be defined only for the TD model, because the 
level of the intrinsic continuum in RD fits is unknown. 
In order to compare the intensity of the soft X--ray power law 
in TD and RD models, we report the monochoromatic flux densities at 1 keV 
in Table~2.

We have also considered an alternative possibility for the 
origin of soft X--ray emission in obscured AGN in terms of 
emission from a photoionized plasma modeled 
in {\tt XSPEC} as {\tt wa$_{Gal}$(pow + N$\times$zgauss + wa(cutoffpl + 2$\times$zgauss))} for TD sources 
and as {\tt wa$_{Gal}$(pow + N$\times$zgauss + pexrav + 2$\times$zgauss))} for RD AGN (Table~3).
$N$ is the  number of fitted lines  in the soft X--ray band which varies form source to source 
(Table~4). The additional {\tt zgauss} component accounts for K$\alpha$ and K$\beta$ 
lines.
Given that the {\tt XIS CCD} energy resolution is not appropriate to fit 
a photoionization code model (Iwasawa et al. 2003), 
the expected soft X--ray spectrum of a photoionized plasma is modeled 
with a power law with photon index free to vary (Table~3), plus intrinsically narrow
(setting $\sigma$=0 in the {\tt zgauss XSPEC} model)  
Gaussian lines at the energy expected (Table 4) for the strongest features observed in high 
resolution {\tt RGS} spectra of obscured AGN (see Guainazzi \& Bianchi 2007).
The presence of the line is evaluated on the basis of a visual 
inspection of the residuals obtained by fitting a single power law.
The best fit energies, fluxes and associated 1$\sigma$ errors of the  
various lines for each source are reported in Table~4.

The best fit spectral parameters of the continuum along with observed X--ray 
fluxes and 2--10 keV intrinsic luminosities are summarized in Table 2 for the 
fits with a partial covering and a thermal spectrum for the soft X--rays, and in Tables 3 and 4 when the 
soft X--rays are modeled with a power law plus Gaussian lines. 
The shape of the hard X--ray continuum is rather independent on the 
model adopted for the soft X--ray spectrum. Therefore fluxes and luminosities 
are reported only in Table 2. 
Galactic absorption using Morrison \& McCammon (1983) cross sections 
(model {\tt wa$_{Gal}$ in XSPEC}) is fixed at the values measured by Dickey \& Lockman (1990).

The detailed results of spectral fits for each source are summarized in the following.

\subsection{NGC 4992} 

The TD  model provides an acceptable fit to the observed spectrum of NGC 4992, 
while the RD model is ruled out at an extremely high confidence level ($\Delta\chi^2 >$ 600).   
The primary continuum is obscured by cold gas, with a column density of 
$N_H \sim 5.5 \times 10^{23}$ \cm\, almost fully covering the central source. 
The iron line intensity ($EW \sim 340$ eV) is consistent with being originated 
by transmission through the obscuring gas.  
The constraints on the fraction of the scattered
($f_{scatt} <$ 0.3 \%) and reflection intensity (R $\simeq$ 0.2--0.5) are rather stringent. 
The soft X--ray emission is extremely weak, and both a thermal spectrum 
($kT \sim$ 0.3 keV) and a steep power law plus
three emission lines at the energies corresponding to the transitions reported 
in Table 4 provide an acceptable fit. 
We note that in the latter case no continuum is formally required, and the entire
soft X--ray flux may be explained as emission from multiple lines.

\subsection{NGC 5728} 

The TD model provides the best fit to the broad band spectrum of NGC 5728. A RD model 
can be safely discarded on a statistical basis ($\Delta\chi^2 >$ 130). The source is mildly 
Compton thick ($N_H \simeq 1.4 \times 10^{24}$ \cm) and thus hard X--rays ($>$ 10 keV) are 
piercing through the obscuring gas. 
The reflection component is rather weak ($R \simeq$ 0.1--0.3). A strong (EW $\sim$ 1 keV) 
iron K$\alpha$ line is detected along with a line at $\sim$ 7 keV. The best fit energy of the 
$\sim$ 7 keV feature, its intensity (EW $\sim$ 70--90 eV) and the observed 
intensity ratio with respect to K$\alpha$, favour an interpretation in terms of 
K$\beta$ emission. The soft X--ray spectrum below a few keV 
is fitted by a combination of a scattered power law, with $f_{scatt} \simeq$ 0.4--1.2 \% ,
and thermal plasma emission ($kT \sim$ 0.3 keV).  The residuals below a few keV suggest 
the presence of line emission. A fit with a steep power law plus several emission lines (Table~4)
also provides a statistically acceptable solution (Table~3) and smooth residuals 
in the soft X--ray band.

\subsection{ESO 263--G13}

The {\it Suzaku} spectrum of ESO263--G13 is best fitted by a TD model. A RD model 
fails to fit the data over the entire energy range ($\Delta\chi^2 >$ 700).
The absorption column density is the 
lowest in the present sample ($\sim 3 \times 10^{23}$ \cm ); the scattering fraction 
and the reflection component intensity are poorly constrained 
($f_{scatt} <$ 0.8 \% and $R \simeq$ 0.1--1.2, respectively). 
The iron K$\alpha$ line EW ($\sim$ 80 eV) is consistent 
with that expected by transmission through the observed column density.
The soft X--ray spectrum is characterized by low counting statistics. 
Both a power law plus thermal X--rays ($kT \sim$ 0.5 keV) 
and a power law plus two emission lines provide an  acceptable fit. 
The latter fit is slightly better in terms of $\chi^2$, though the improvement 
is not statistically significant.

\subsection{ESO 137--G34}

Both a RD model ($\chi^2/d.o.f \simeq$ 169/139 see Table 3) and a TD model 
($\chi^2/d.o.f \simeq$ 160/138) provide an acceptable fit to the broad band spectrum
of ESO 137--G34. In the TD fit, the primary hard X--ray continuum ($\Gamma \simeq$ 1.8), 
is obscured by a column density of $\sim 1.2 \times 10^{24}$ cm$^{-2}$. The intensity of the 
scattering fraction ($f_{scatt} \simeq$ 5\%) is higher than that measured in the 
sources best fitted with a TD model, while the reflection component is relatively 
weak ($R \sim$ 0.3). Even though none of the two models cannot be preferred on the 
basis of purely statistical arguments, the individual parameters of the TD model 
are highly degenerate, and cannot be reliably constrained.
As a consequence, a RD model is preferred. It is worth noting that the source 
would be classified as Compton Thick in both cases. 
The strong K$\alpha$ line (EW $\sim$ 1.5 keV) is accompanied by  a line at 
$\sim$ 7 keV. The best--fit line intensity ratio ($\sim$ 0.2) is slightly higher, but not
inconsistent within the errors, than that expected by a K$\beta$ line origin for 
the $\sim$ 7 keV feature.
The soft X--ray spectrum can be fitted with a thermal plasma ($kT \sim$ 0.8 keV) 
and a power law slope which is significantly steeper than that of the hard X--ray continuum.  
Some residual line like emission, 
around  1 keV and 1.8 keV, is still present, suggesting a more complex spectrum.
A fit with a power law plus several Gaussian lines  (Table~4) provides an  acceptable solution
in terms of $\chi^2$ statistic and smooth residuals (see Fig.~4).

\subsection{ESO 323--G32} 

Both a RD model ($\chi^2/d.o.f \simeq$ 79/89 see Table 3) and a TD model 
($\chi^2/d.o.f \simeq$ 83/88)  provide an acceptable fit to the broad band spectrum.
In the TD fit, the primary hard X--ray continuum $\Gamma \simeq$ 2.0, is obscured 
by a column density  of the order of $\sim 1-2 \times 10^{24}$ cm$^{-2}$. The relative intensity 
of the transmitted component is about a factor two lower than that of the reflection component.
Similarly to ESO 137--G34, the best fit spectral parameters, 
in particular the column density $N_H$ and the scattering fraction, cannot be constrained.
A RD model is thus preferred as the best fit to ESO 323--G32, the faintest source of the sample. 
The iron K$\alpha$ line is very strong (EW $\sim$ 2 keV), and the best fit energy and 
intensity of the $\sim$ 7 keV feature are consistent with a K$\beta$ origin.
The weak soft X--ray emission can be adequately modeled with 
either a power law plus a thermal spectrum or with a power law plus 
two Gaussian lines. Both fits are statistically acceptable and comparable.

\subsection{Summary of fitting results}

\subsubsection{Hard X--ray spectra}

All the targets are obscured by column densities larger than $\sim 3 \times 10^{23}$ \cm.
Two of them are highly absorbed but Compton thin ($N_H \simeq 3-5 \times 10^{23}$ cm$^{-2}$), 
while the other three are Compton Thick. 
In NGC 5728 the primary hard X--ray continuum is transmitted through a column density 
$N_H \simeq 10^{24}$ cm$^{-2}$. In ESO 137--G34 and ESO 323--G32,
both a TD and a RD model provide an acceptable description 
of the broad band spectra. An interpretation in terms of a reflection dominated 
continuum is preferred and considered more reasonable (see $\S 4.4$ and $\S 4.5$).
The photon indices are in the range 1.4--1.9 and  thus typical, or possibly 
slightly flatter than the average values of Seyfert 1 galaxies.
This suggests that the intrinsic continuum, when visible, is reasonably well constrained,
and highlights the importance of the hard X--ray band in the study of heavily obscured
AGN. 
The absorption corrected luminosities of the sources best fitted by the TD model are in the
Seyfert range ($L_{2-10\  keV} \sim 10^{42-43}$ \lum). The observed hard X--ray luminosities 
of the reflection dominated AGN are of the order of 10$^{41}$ \lum. 
The intrinsic luminosity can be approximated by $L_{int} \sim L_{obs}/(C \times A_{2-10}$) 
where $C = \Delta\Omega/4\pi$ and $\Delta\Omega$ is the solid angle illuminated by the central
source reflecting X--rays, $A_{2-10}$ is the albedo of Compton reflection 
in the 2--10 keV energy range. The albedo is a weak function of the slope of the 
primary spectrum and, for the range of slopes of the sources in our sample, 
is of the order of 7\%, assuming reflection from a slab covering a solid angle 
of $2\pi$ at the source. We note that the albedo due to torus reflection is likely 
to be lower than the quoted value (Murphy \& Yaqoob 2009). 
The intrinsic luminosities
for the two reflection dominated Seyferts are 1.4 -- 3.2 $\times 10^{42} C^{-1}$ \lum ,
consistent with the X--ray luminosities of Seyfert galaxies. 
The ratio between observed and intrinsic luminosity 
in the Circinus galaxy (Matt et al. 1999) is about a factor two lower than that estimated 
for the objects in our sample. Given that C $<$ 1 and considering the 
uncertainties associated to the Compton reflection albedo from distant matter,  
the agreement is rather good.
The reflection component intensity, for those sources best fitted with a TD model, 
is poorly constrained; however, it seems to be relatively weak. 
The iron K$\alpha$ line equivalent widths in Compton thin sources are consistent with a 
transmission origin through the observed column densities. In the RD sources, the 
large EW are further supporting the presence of a reflection dominated continuum 
in these objects.
An emission line feature at $\approx$ 7 keV (rest frame) is detected in the 
most obscured AGN. The best fit line energy and measured intensity are consistent,
in all cases, with iron K$\beta$ emission.

\subsubsection{Soft X--ray spectra} 

Soft X--ray emission is ubiquitous in the sources of our sample, although the intensity
and shape of the soft X--ray spectrum vary significantly from object to object (Fig.~1).
The fraction of scattered flux, in the three objects whose hard X--ray continuum 
is best fitted with a TD model, is relatively low and below about 1\%.
These sources would be dubbed  ``hidden" AGN by Winter et al. (2008).
In all sources, including the two RD objects for which the scattering 
fraction can not be defined, a soft thermal component with temperatures in the range 
0.3--0.8 keV is also required. The 0.2--2 keV X--ray luminosity associated to the 
thermal component is in the range 1--7 $\times 10^{40}$ \lum , 
consistent with the values observed in ``normal" galaxies (i.e. Norman et al. 2004).
While the physical origin of the thermal component could be associated to 
diffuse hot plasma in the host galaxy, it is important to note that the best fit
temperatures are derived by fitting the iron $L$--shell emission at 0.7--1 keV. 
We also note that line--like features below $\sim$ 2 keV are present in 
the residuals, with respect to a fit with a thermal component and a power law (Fig.~3).
Prompted by these considerations, we explored a different possibility 
for the origin of soft X--ray emission.
A photoionized plasma is found to be an excellent description of the 
soft X--ray spectra for a large sample of obscured Seyfert 2 
observed with the high resolution {\sc rgs} gratings on board 
XMM (Guainazzi \& Bianchi 2007).
The spectrum of photoionized gas is usually modeled with specific 
codes (i.e. {\sc xstar}). However, the available counting statistics and the  
limited {\sc ccd} energy resolution of the present data are not such to adequately 
constrain the physical properties of the emitting plasma.
We adopted a simplified approach fitting the observed spectra 
with a power law plus individual narrow emission lines.
The best fit line energy and intensities are reported in Table~4.
In two sources of our sample, NGC 5728 and ESO 137-G34  (both Compton thick),
the parameterization described above provides a formally 
better description of the soft X--ray spectra, though not statistically significant, 
than a power law plus thermal emission.
A comparison of the best fit {\it Suzaku} spectra of three objects in the sample 
is shown in Fig.~3.
From the analysis of the residuals, it is clear that line--like emission 
below 2 keV is not properly accounted for by a thermal plasma fit.

\section{Discussion and Conclusions}

Thanks to the broad band energy range and sensitivity of the X--ray detectors onboard 
{\it Suzaku} it has been possible to study, with good accuracy, 
the spectral properties of a sample of five hard X--ray selected type 2 AGN.
The presence of cold obscuring gas has been unambiguosly established 
in all the sources. Although this result was expected, given the sample selection criteria, 
the X--ray spectra show a high degree of spectral complexity. In particular, the 
absorption column density and the relative intensity of the various 
spectral components differ from object to object.

The primary emission is seen, in three out of five sources, through 
column densities in the range $\sim 10^{23.5-24}$ \cm. The fraction of 
the intrinsic flux scattered into the line of sight and parameterized 
by $f_{scatt}$ in the partial covering fits is relatively low and the reflection 
component intensity weak. 
On the basis of {\it Suzaku} observations of six {\it Swift} selected AGN,
thus an almost identical selection criterium,  Eguchi et al. (2009) found 
that sources with a low scattering fraction ($f_{scatt} <$  0.5 \%) also have a relatively 
strong reflection component ($R$ $\gsimeq$ 1). They dubbed these objects as 
``new" type AGN (see also Ueda et al. 2007) and proposed they may represent 
the tip of the iceberg of a hitherto uncovered population of AGN obscured 
by a geometrically and optically thick torus (Levenson et al. 2002). 
\pn
The analysis of the present data does not support the ``new" type AGN interpretation. 
A possible explanation may be due to the strong degeneracy in the fit parameters.
For example, if  the reflection component is also absorbed (like in the Eguchi et al. 2009), 
a solution in terms of a stronger reflection intensity and lower normalization 
of the scattered component is found, and results similar to Eguchi et al. (2009)
are recovered also in our sample (Comastri et al. 2009). 
However, we believe that the quality of our data does not allow us to constrain 
absorption on the reflection component; moreover, 
our best fit solution has the advantage to have a lower number of free 
parameters.
It is also worth pointing out that the intensity of the reflection spectrum 
from Compton thick gas strongly depends on the geometry of the system.
The reflection fraction, obtained from fitting  disc--reflection models to 
Compton thick AGN, cannot be directly associated to a solid angle (Murphy \& Yaqoob 2009). 
A better understanding of the nature of ``new" type AGN and whether there are
two distinct classes of Seyfert 2, possibly related to the geometry of the 
obscuring medium, or a continuous distribution of spectral parameters requires
a larger sample of objects with a good counting statistic to allow for a detailed 
spectral analysis.
\pn
The observational evidence of obscured AGN with a low 
scattering fraction is at variance 
with the average values adopted for the above parameters in the Gilli et al. (2007)
XRB synthesis models.
We have tested the impact of different normalizations for the scattering component in absorbed 
($N_H > 10^{22}$ \cm ) AGN. The fit to the XRB spectrum and source counts 
are basically unaffected, unless $f_{scatt}$ exceeds values of the order of 10 \%.
High $f_{scatt}$ values appears to be the exception, rather than the rule, 
and are definitely ruled out among highly obscured {\it Swift} selected AGN. 
\pn 
The possibility that photoionized gas is responsible for the soft X--ray emission 
in obscured AGN is consistent with the present observations. 
It is interesting to note that 
the residuals in the soft X--ray band of the NGC 5728 {\it Chandra} 
spectrum (Zhang et al. 2006) and ESO 137--G34 XMM--{\it Newton} spectrum (Malizia et al. 2009), 
both fitted with thermal emission, suggest the presence of unaccounted for 
emission lines at energies close to those reported here.
Unfortunately, the physical status of the ionized gas cannot be constrained 
by the present observations. Only a limited number of emission lines are individually 
detected above the 3$\sigma$ level. It is however reassuring that 
the best fit energies reported in Table~4
are consistent with those expected by ionised metals observed with the 
XMM--{\it Newton} and {\it Chandra} gratings observations of much brighter 
Seyfert 2 galaxies (Brinkman et al. 2002; Kinkhabwala et al. 2002; Guainazzi \& Bianchi 2007).
The $\sim$ 0.9 keV feature, identified with {\sc Ne ix} appears
to be ubiquitous, possibly with the exception of NGC 4992 which 
is, however, the faintest soft X--ray source in the sample. 

It is worth noting that the three brightest soft X--ray sources in the sample, 
all of them Compton thick (Fig.~1), show strong {\sc [oiii]} ionization cones 
in optical images (e.g. Wilson et al. 1993). Moreover, extended soft X--ray emission 
almost spatially coincident with {\sc [oiii]} cones is detected 
by {\it Chandra} imaging observations of NGC 5728 (Zheng et al. 2006). 
Furthermore, ESO 137--G34 and ESO 323--G32 are hosted by S0 galaxies which 
are expected to have neither intense star formation activity, nor thermal X--ray emission 
from hot gas. 

Although it may be premature to discard, on the basis of the present X--ray 
observations, a thermal origin for the soft X--ray emission, we stress that the 
above described arguments would support the presence of photoionized plasma,
which is also favored by high resolution {\sc rgs} observations of 
bright obscured AGN (Guainazzi \& Bianchi 2007).
\pn 
Long--look observations of selected sample of obscured AGN are clearly needed 
to elucidate the nature of the soft X--ray emission in these objects, 
though firm conclusions can be drawn only by high resolution 
(with either gratings or calorimeters) X--ray observations.

\acknowledgments
We thank Giovanni Miniutti for extremely useful discussions. 
Support from the Italian Space Agency (ASI) under the 
contracts ASI-INAF I/088/06/0 and ASI-INAF I/23/05 
is acknowledged.

\clearpage

\begin{deluxetable}{ccccc}
\tabletypesize{\scriptsize}
\rotate
\tablecaption{The Sample} 
\tablewidth{0pt}
\tablehead{
\colhead{Name} & $z$ & Exposure Time (ks) & {\sc xis}$^a$  & {\sc pin}  
}
\startdata
NGC 4992     & 0.025   & 31.5  &  3.30$\pm$0.06  & 6.4$\pm$0.5    \\
NGC 5728     & 0.009   & 37.1  &  3.10$\pm$0.06  & 7.9$\pm$0.4    \\
ESO263-G13   & 0.033   & 37.4  &  6.30$\pm$0.09  & 4.0$\pm$0.4    \\
ESO137-G34   & 0.009   & 78.0  &  1.80$\pm$0.04  & 1.3$\pm$0.2    \\
ESO323-G32   & 0.016   & 80.5  &  1.00$\pm$0.04  & 0.7$\pm$0.2    \\ 

\enddata
\tablecomments{Background subtracted count rates in units of 10$^{-2}$ counts s$^{-1}$.}
\tablenotetext{a}{XIS count rates are averaged over 2 or 3 CCD units.}
\end{deluxetable}

\clearpage

\begin{deluxetable}{ccccccc}
\tabletypesize{\scriptsize}
\rotate
\tablecaption{Best fit spectral parameters}
\tablewidth{0pt}
\tablehead{
\colhead{Parameters} & \colhead{NGC 4992} & \colhead{NGC 5728} & \colhead{ESO263-G13} & 
\colhead{ESO137-G34} & \colhead{ESO323-G32} & 
}
\startdata
$N^{Gal}_H$ (10$^{22}$ \cm) & 0.02       & 0.07         & 0.11             & 0.24 & 0.08  \\
$N_H$ (10$^{22}$ \cm)       & 56$\pm$6   &  139$\pm$15     & 28$\pm$3     & 1000(a)   &  1000(a)  \\
$\Gamma_H$                    & 1.56$\pm$0.13  & 1.70$\pm$0.16  & 1.91$^{+0.19}_{-0.13}$  & 1.43$^{+0.24}_{-0.16}$ & 1.62$\pm$0.45 \\
$f_{scatt}$  (\%)           & $<$0.3 & 0.7$^{+0.5}_{-0.3}$    & $<$0.8   &  --$^b$   &  -- \\
$f_{1 keV}^c$ 		    & $<$0.63 & 3.3 & $<$2.9 & 4.3 & 1.9  \\
$R$ 	                    & 0.3$^{+0.2}_{-0.1}$  & 0.2$\pm$0.1  & 0.5$^{+0.7}_{-0.4}$  & $\infty$ (a)    & $\infty$(a) \\
$E(K\alpha)$ (keV)           & 6.38$^{+0.02}_{-0.03}$ & 6.39$\pm$0.01 & 6.39$\pm$0.04 & 6.40$\pm$0.01 & 6.42$\pm$0.02  \\
$EW^c$ (keV)                   & 0.34$\pm$0.07 & 1.03$^{+0.39}_{-0.40}$ & 0.08$\pm$0.04 & 1.46$^{+0.16}_{-0.22}$ & 1.93$\pm$0.28 \\
$E(K\beta)$ (keV)            & -- & 7.05$\pm$0.04 & -- & 6.98$^{+0.05}_{-0.07}$  & 6.97$^{+0.08}_{-0.06}$    \\
$EW^d$ (keV)                 & -- & 0.07$^{+0.05}_{-0.04}$ & -- & 0.31$^{+0.12}_{-0.13}$ & 0.26$\pm0.15$     \\
$kT$ (keV)                   & 0.34$^{+0.35}_{-0.13}$ & 0.28$\pm$0.03 & 0.52$^{+0.08}_{-0.24}$ & 0.84$\pm$0.06  & 0.78$\pm$0.20  \\
$F$ (0.5-2 keV)              & 0.02 & 0.16 & 0.07 & 0.14 & 0.05    \\
$F$ (2-10 keV)               & 2.4 & 1.5 &  3.8 & 0.53 & 0.40   \\
$L$ (2-10 keV)               & 1.6 & 0.48 & 2.7 & 0.01(e) & 0.023(e)    \\
$\chi^2/d.o.f.$              & 126.0/123 & 169.5/143 & 177.8/207 & 169.7/139  & 79.0/89  \\

\enddata
\tablecomments{ Fluxes are in unit of $10^{-12}$ \cgs, absorption corrected luminosities 
in units of $10^{43}$ \lum. }
\tablenotetext{a}{For the sources best fitted with a RD model, the 
column density is fixed to $10^{25}$ \cm  and the ratio of the reflected to the transmitted 
component diverges by definition.}
\tablenotetext{b}{No acceptable solutions are found linking the two power law indices.
Leaving both the power law slopes free to vary, the best fit in the soft X--rays is 2.45$\pm$0.40.}
\tablenotetext{c}{Fluxes at 1 keV in units of 10$^{-5}$ ph cm$^{-2}$ s$^{-1}$ keV$^{-1}$.} 
\tablenotetext{d}{The lines EW is computed against the total continuum at the best fit line energy. For the sources
best fitted by a TD model, the line is also absorbed by the quoted column density.} 
\tablenotetext{e}{Observed luminosities.}
\end{deluxetable}

\clearpage

\begin{deluxetable}{ccccccc}
\tabletypesize{\scriptsize}
\rotate
\tablecaption{Best fit spectral parameters}
\tablewidth{0pt}
\tablehead{
\colhead{Parameters} & \colhead{NGC 4992} & \colhead{NGC 5728} & \colhead{ESO263-G13} & 
\colhead{ESO137-G34} & \colhead{ESO323-G32} 
}
\startdata
$N^{Gal}_H$ (10$^{22}$ \cm)   & 0.02            &  0.07                   & 0.11                   & 0.24 & 0.08 \\
$N_H$ (10$^{22}$ \cm)         & 55$\pm$6         &  140$^{+5}_{-8}$          & 26$\pm$1           & 1000(a)   &  1000(a)  \\
$\Gamma_H$                    & 1.56$^{+0.11}_{-0.14}$ &  1.68$^{+0.05}_{-0.04}$   & 1.79$^{+0.05}_{-0.04}$ & 1.58$^{+0.20}_{-0.16}$ & 1.85$^{+0.47}_{-0.40}$  \\
$\Gamma_S^b$                    & ...(c) & 2.7$\pm$0.3   & 2.40$\pm$0.90  & 3.11$^{+0.28}_{-0.24}$ & ...(d)      \\
$f_{1 keV}^e$ 		      & $<$0.83 & 6.3 & 3.1 & 12.6 & 2.3  \\
$R$ 	                      & 0.32$\pm$0.12  & 0.34$^{+0.04}_{-0.08}$  & $<$ 0.68   & $\infty$ (a)    & $\infty$(a)    \\
$E(K\alpha)$ (keV)            & 6.38$^{+0.02}_{-0.03}$ & 6.39$\pm$0.01 & 6.39$\pm$0.05 & 6.40$\pm$0.01 & 6.42$\pm$0.02    \\
E.W. (keV)                    & 0.35$\pm$0.07 & 1.28$^{+0.14}_{-0.13}$ & 0.08$^{+0.045}_{-0.035}$ & 1.6$\pm$0.2 & 2.0$\pm$0.3 \\
$E(K\beta)$ (keV)             & -- & 7.05$\pm$0.05 & -- & 6.99$^{+0.06}_{-0.08}$  & 6.97$\pm$0.08    \\
$E.W.$ (keV)                  & -- & 0.09$^{+0.07}_{-0.06}$ & -- & 0.35$^{+0.12}_{-0.20}$ & 0.27$\pm$0.17     \\
$\chi^2/d.o.f.$               & 120.3/119 & 135.3/133 & 172.7/206 & 146.7/131  & 78.8/87   \\

\enddata
\tablecomments{Spectral fits parameters obtained when the soft X--ray spectrum is fitted 
with a power law plus individual Gaussian lines.}
\tablenotetext{a}{For the sources best fitted with a RD model, the column density is fixed to $10^{25}$ \cm and 
the ratio of the reflected to the transmitted 
component diverges by definition.}
\tablenotetext{b}{Photon index of the power law representing the photoionized continuum.}
\tablenotetext{c}{The continuum flux is extremely weak and formally consistent with zero.}
\tablenotetext{d}{The power law slope in the soft band is unconstrained and fixed to be equal to the hard power law slope.}
\tablenotetext{e}{Fluxes at 1 keV in units of 10$^{-5}$ ph cm$^{-2}$ s$^{-1}$ keV$^{-1}$.} 
\end{deluxetable}

\clearpage

\begin{deluxetable}{ccccccc}
\tabletypesize{\scriptsize}
\rotate
\tablecaption{Soft X-ray lines} 
\tablewidth{0pt}
\tablehead{
\colhead{Parameters} & \colhead{NGC 4992} & \colhead{NGC 5728} & \colhead{ESO263-G13} & 
\colhead{ESO137-G34} & \colhead{ESO323-G32} & \colhead{Line ID$^a$}
}
\startdata
$E_1$ (keV)    & 0.70$^{+0.06}_{-0.02}$  &  0.65$\pm$0.02  & 0.69$\pm$0.03  & ...  & ... &   {\sc O viii} (0.65) - {\sc Fe xvii} (0.72) \\
Flux$_1$       & 0.9$\pm$0.5  &  3.7$\pm$1.0    & 1.9$^{+2.2}_{-0.9}$  & ...  & ... &    \\
$E_2$ (keV)    & 0.84$\pm$0.02  &  0.88$\pm$0.01  & 0.87$\pm$0.02  & 0.91$\pm$0.02 & 0.94$\pm$0.02  &  {\sc Ne ix} (0.90-0.92) - {\sc Fe xviii} (0.86)  \\
Flux$_2$       & 0.4$\pm$0.2   &  1.0$\pm$0.2    & 0.94$\pm$0.3   & 0.67$\pm$0.30 & 0.3$\pm$0.1 &   \\
$E_3$ (keV)    &  ...   &  0.98$\pm$0.02  &  ... & 1.03$\pm$0.01 & ...  &  {\sc Ne x} (1.02)  \\
Flux$_3$       & ...   &  0.4$\pm$0.1 & ...    & 0.90$\pm$0.20 & ... &   \\
$E_4$ (keV)    & 1.69$^{+0.06}_{-0.03}$  &  1.76$\pm$0.02  &  ...   &   1.79$\pm$0.02  &  1.81$\pm$0.03   &  {\sc Si xiii} (1.84)  \\
Flux$_4$       & 0.09$\pm$0.04  &  0.11$\pm$0.04  & ... & 0.18$\pm$0.05 & 0.08$\pm$0.04 &   \\
$E_5$ (keV)    & ... &  2.35$\pm$0.03  & ...  & 2.53$\pm$0.05         &  ...     &     {\sc S xv} (2.44)   \\
Flux$_5$       & ... &  0.15$\pm$0.06  & ...  & 0.12$^{+0.03}_{-0.06}$ & ... &   \\
$E_6$ (keV)    & ... &  3.13$\pm$0.04  & ...  & 3.11$\pm$0.02         &  ... &  {\sc Ar xvii}  (3.11) \\
Flux$_6$       & ... &  0.11$\pm$0.05  & ...  & 0.10$\pm$0.04         & ... &   \\
\enddata
\tablecomments{Emission-line fluxes in units of 10$^{-5}$ phot. cm$^{-2}$ s$^{-1}$. 
Errors on both line energy and flux are at 1$\sigma$.}
\tablenotetext{a}{Suggested line identification and corresponding energies in keV (from Guainazzi \& Bianchi 2007 and 
House 1969).
The approximate energy interval for the {\sc Ne ix} triplet is reported.}
\end{deluxetable}

\clearpage

\begin{figure*}[t]
\includegraphics[angle=-90,scale=1]{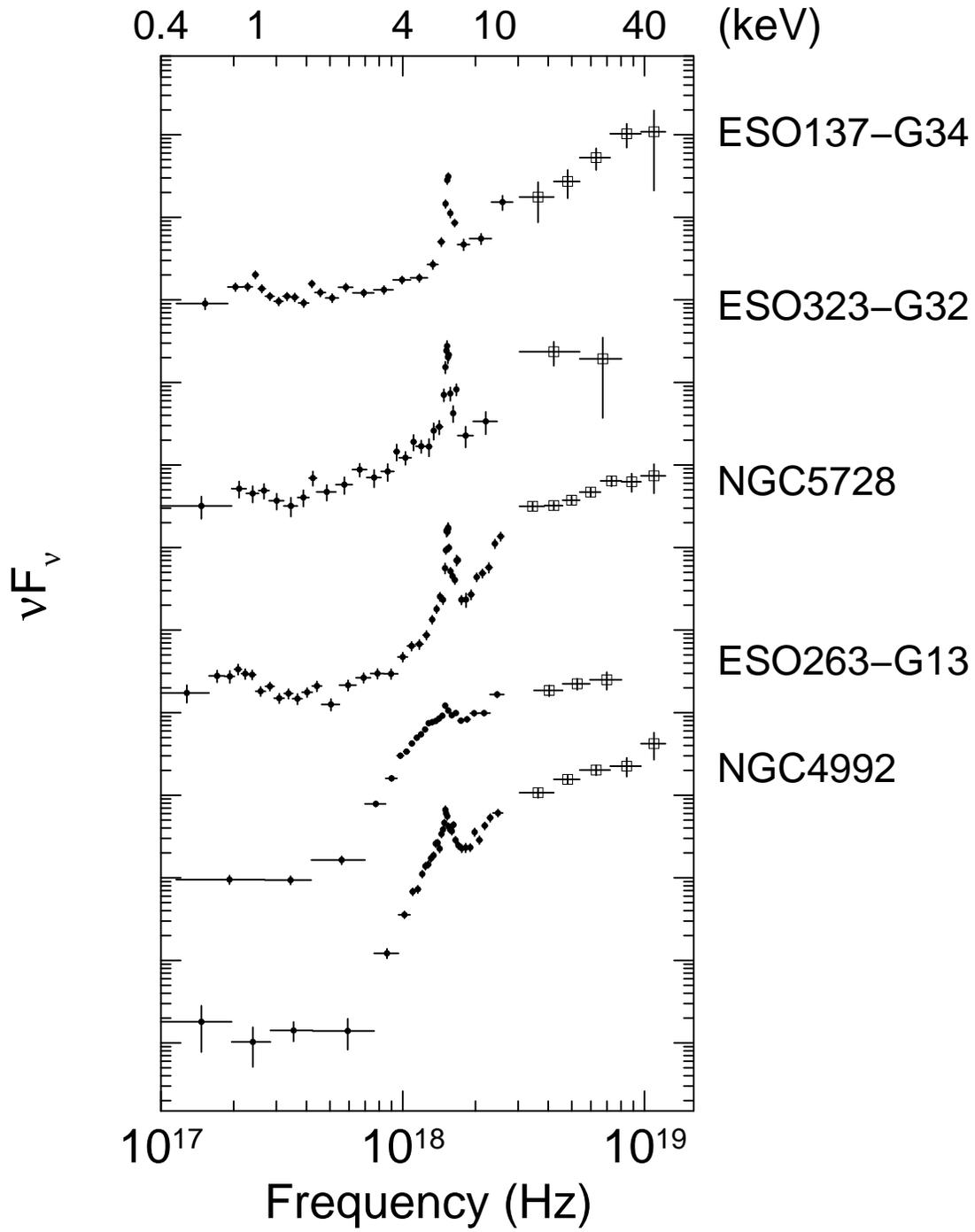}
\caption{The broad band {\it Suzaku} spectra of the five sources, 
covering about two decades in frequency, 
in a $\nu$F$_{\nu}$ representation. The Y-axis normalization is arbitrary.}
\end{figure*}

\clearpage

\begin{figure*}[t]
\includegraphics[keepaspectratio=false, height=7.5cm, width=5.5cm,
angle=270]{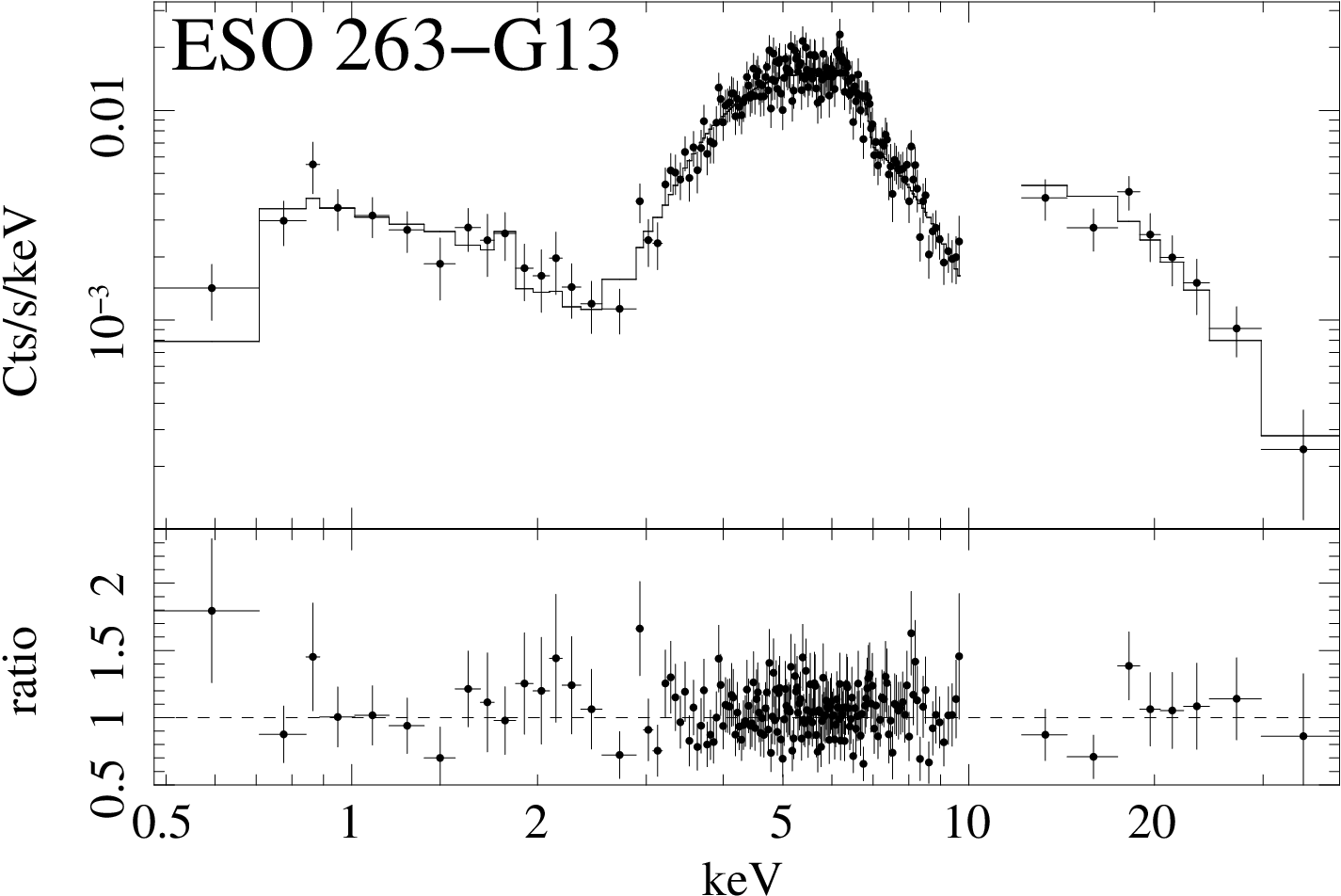}
\includegraphics[keepaspectratio=false, height=7.5cm, width=5.5cm,
angle=270]{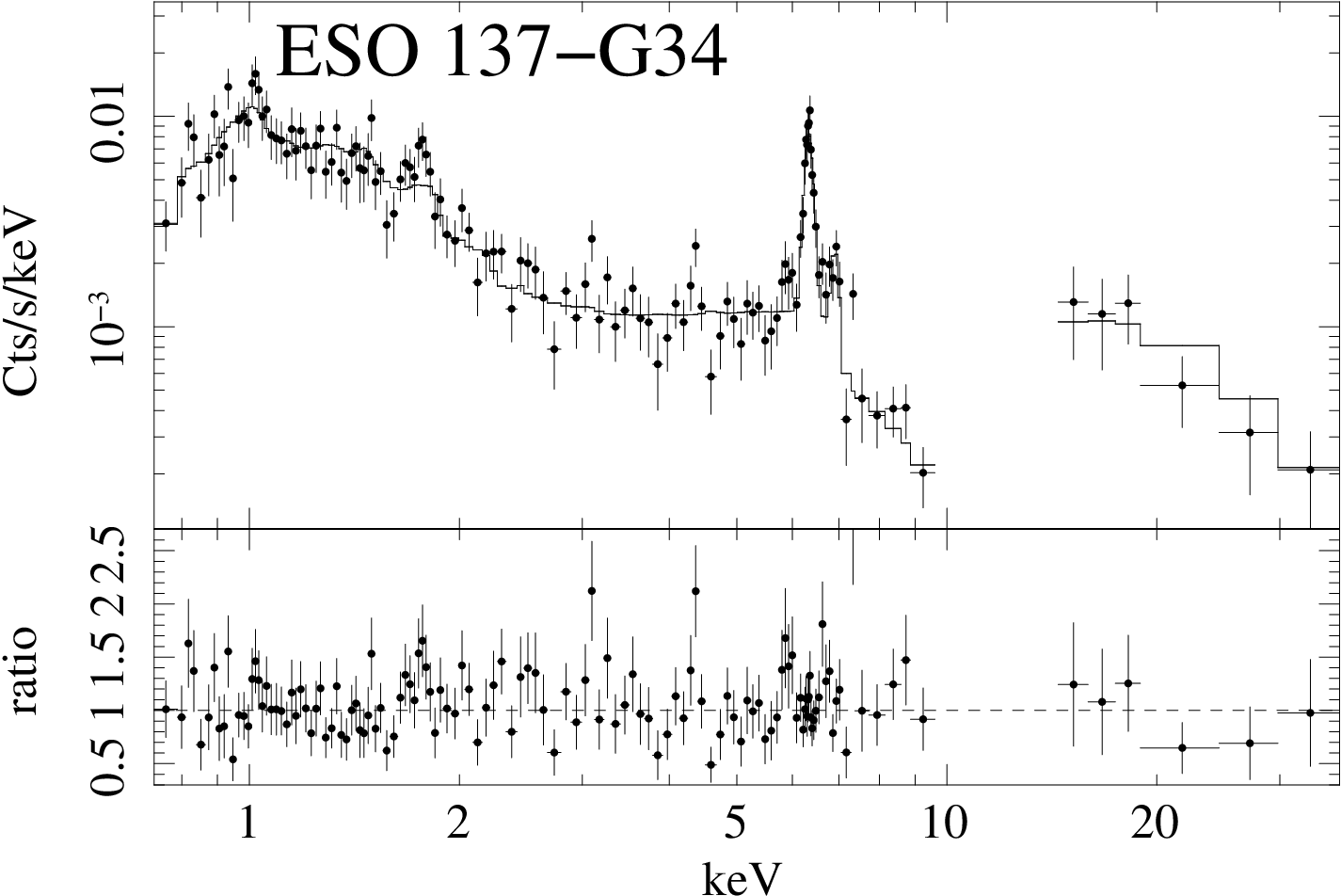}
\includegraphics[keepaspectratio=false, height=7.5cm, width=5.5cm,
angle=270]{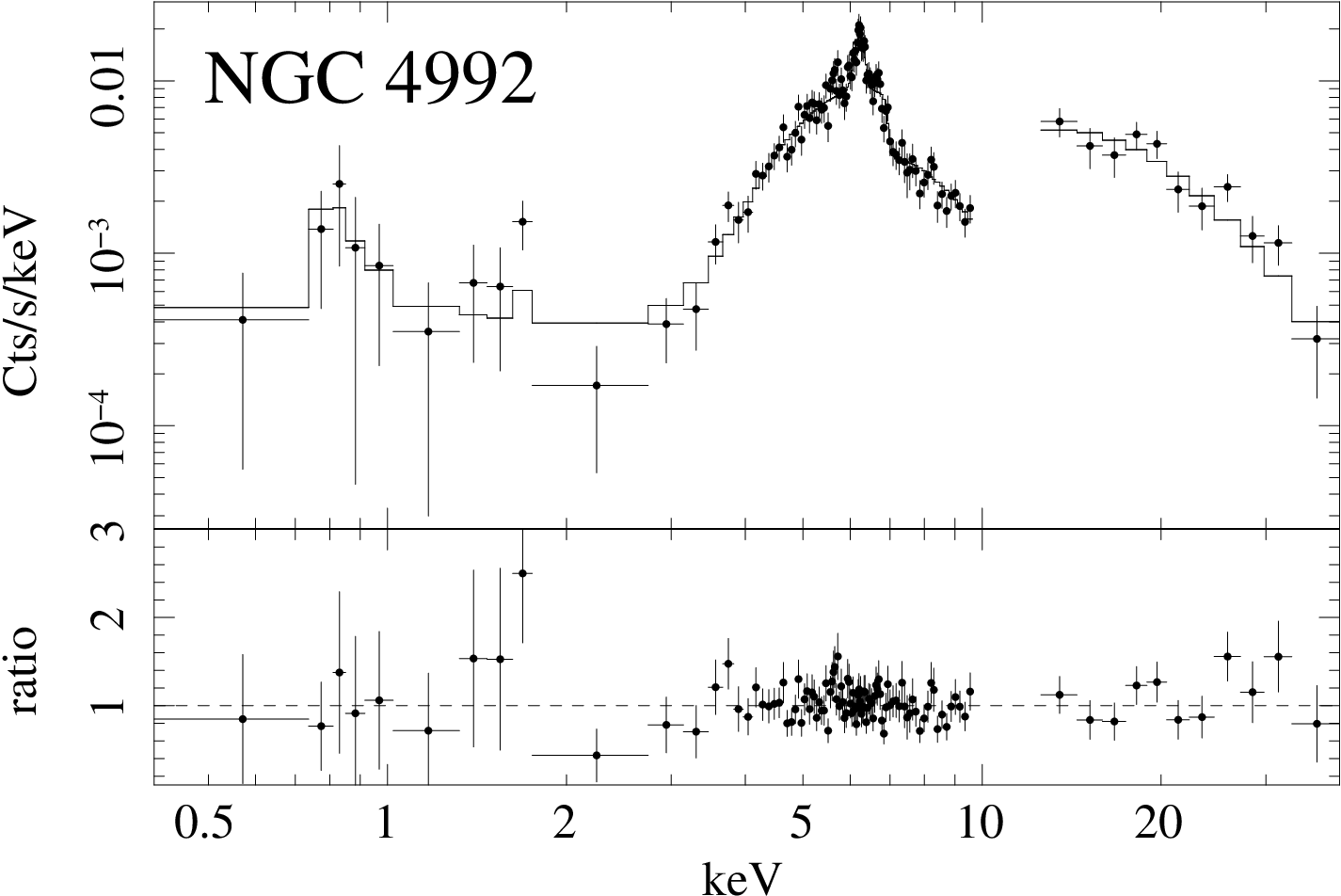}
\includegraphics[keepaspectratio=false, height=7.5cm, width=5.5cm,
angle=270]{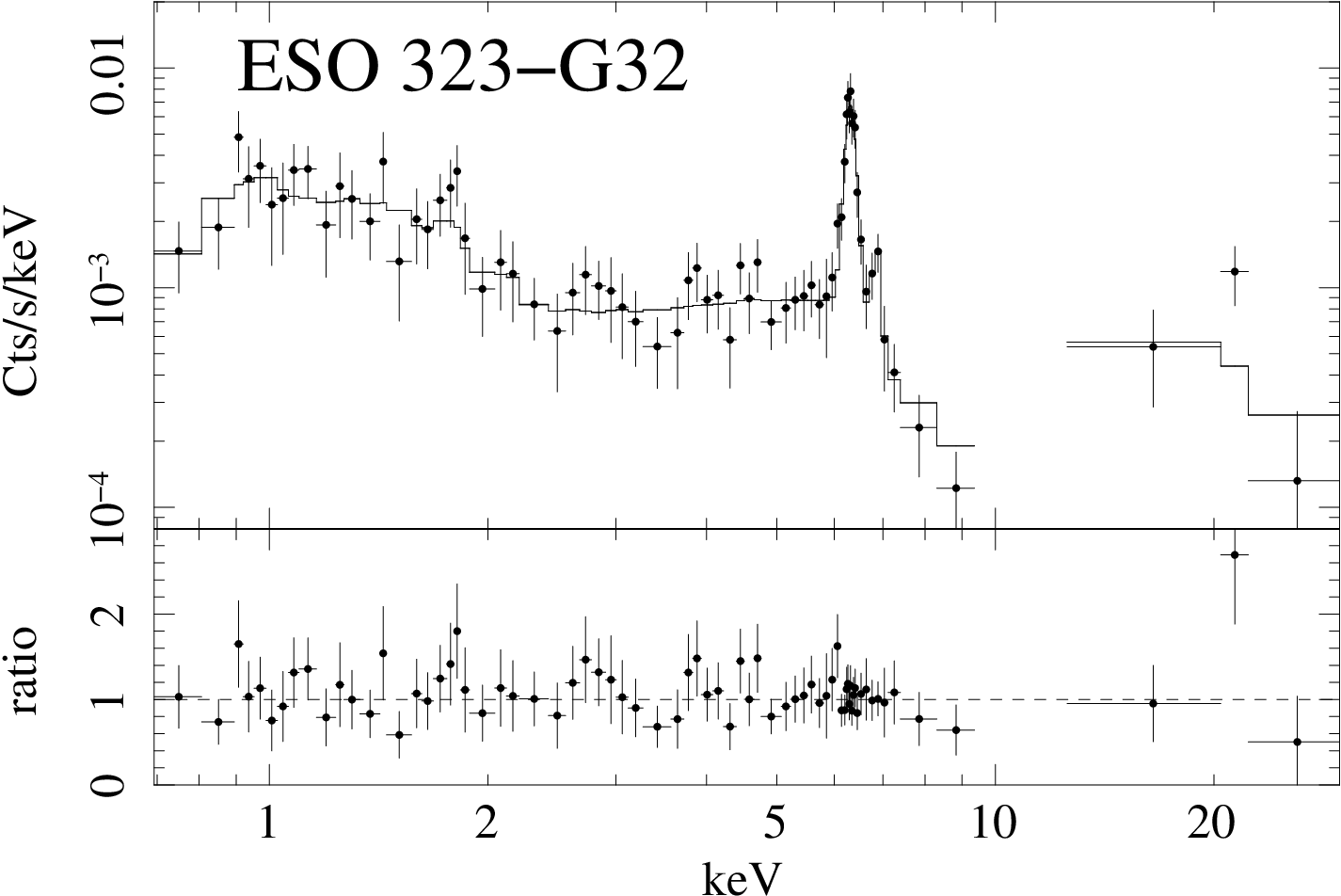}
\includegraphics[keepaspectratio=false, height=7.5cm, width=5.5cm,
angle=270]{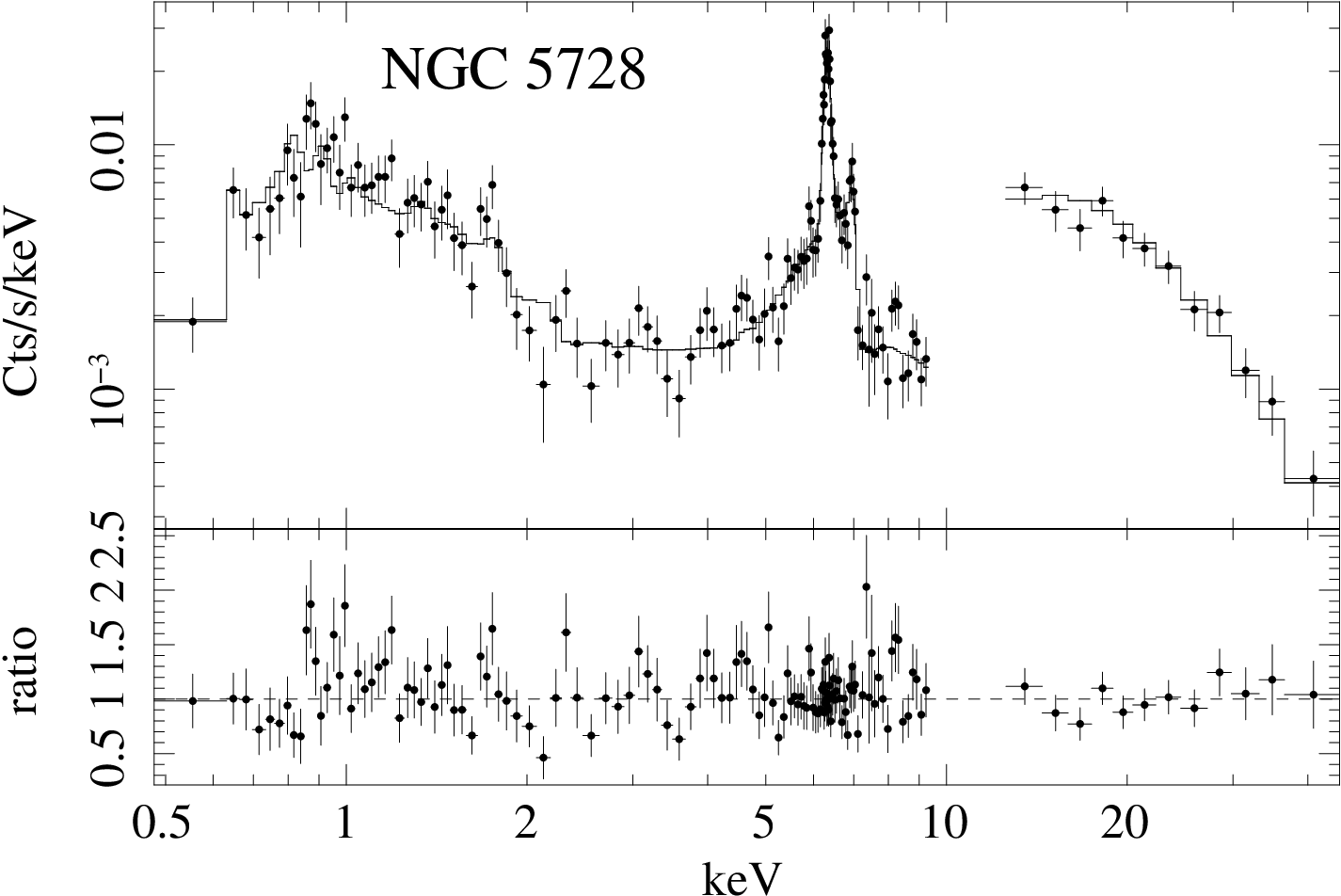}
\caption{The {\it Suzaku} broad band spectra and data to model ratio for the 5 Seyfert galaxies of the sample.
In the left column we plot the sources best fitted with a TD model,
and in the right column the two reflection dominated X--ray spectra. The spectral 
parameters of the broad band fits shown in the figures are reported in Table 2.}
\label{oxspec}
\end{figure*}

\clearpage

\begin{figure*}[t]
\begin{center}
\includegraphics[keepaspectratio=false, height=7.5cm, width=5.5cm,
angle=270]{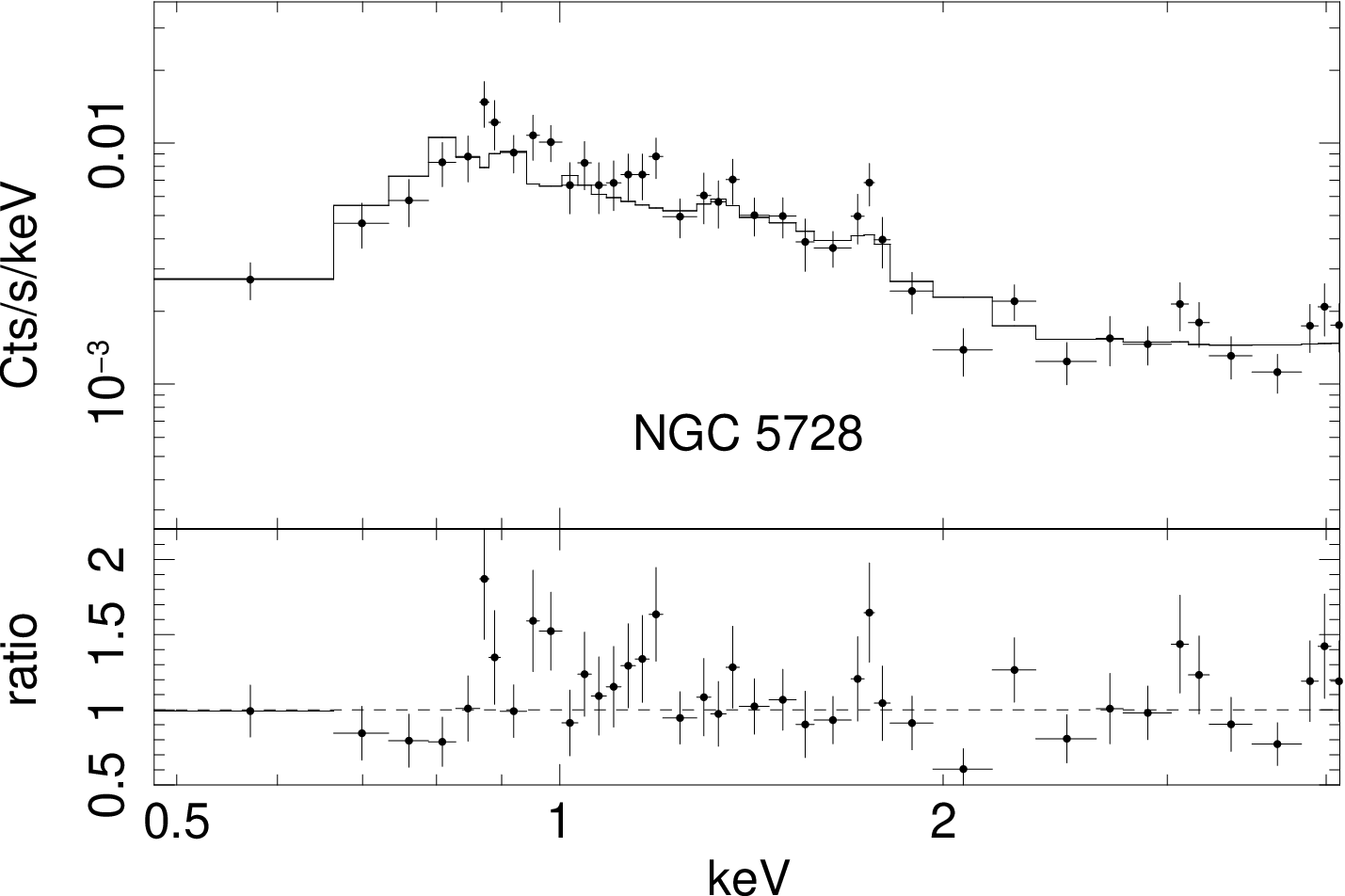}
\includegraphics[keepaspectratio=false, height=7.5cm, width=5.5cm,
angle=270]{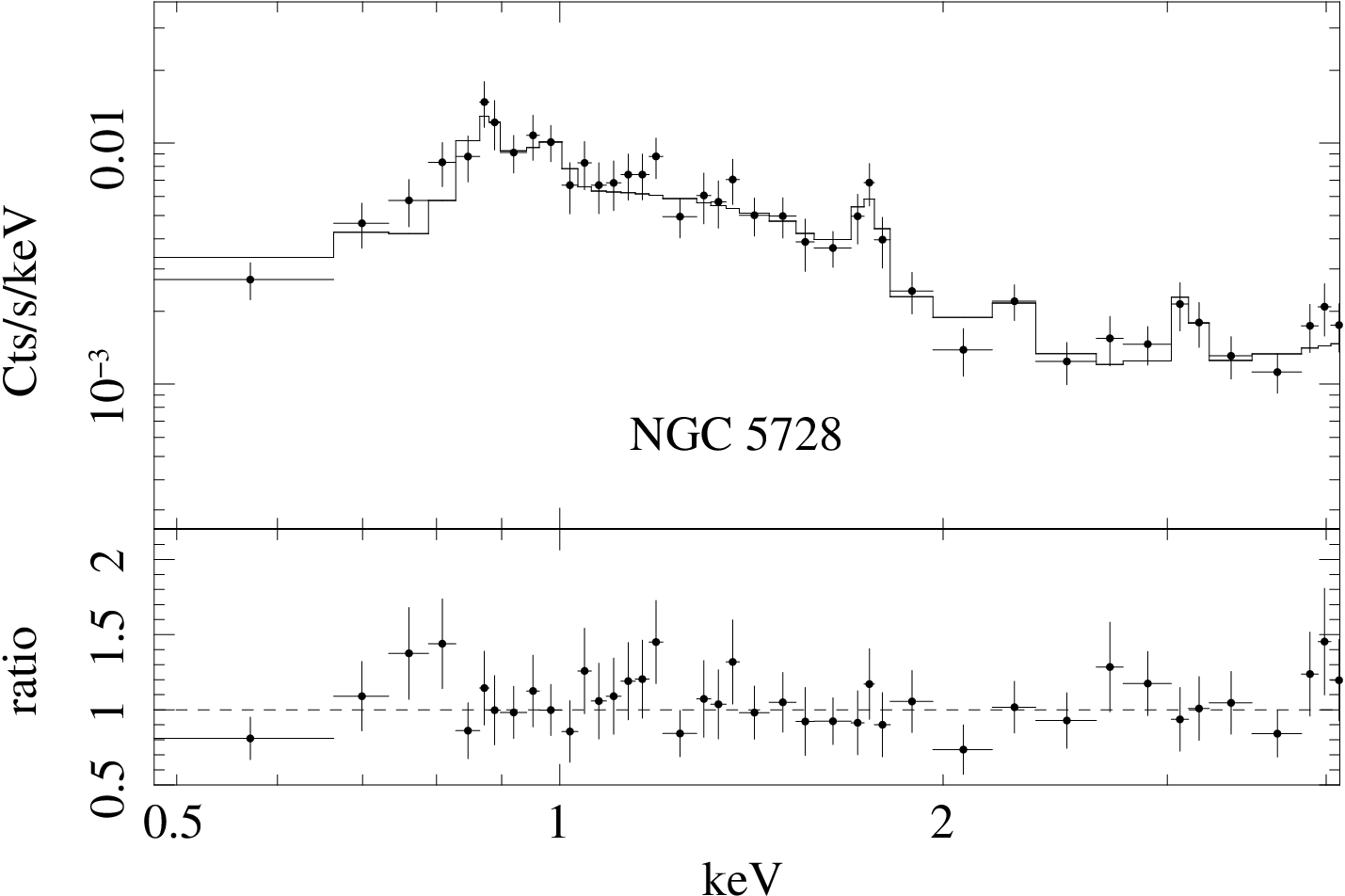}
\includegraphics[keepaspectratio=false, height=7.5cm, width=5.5cm,
angle=270]{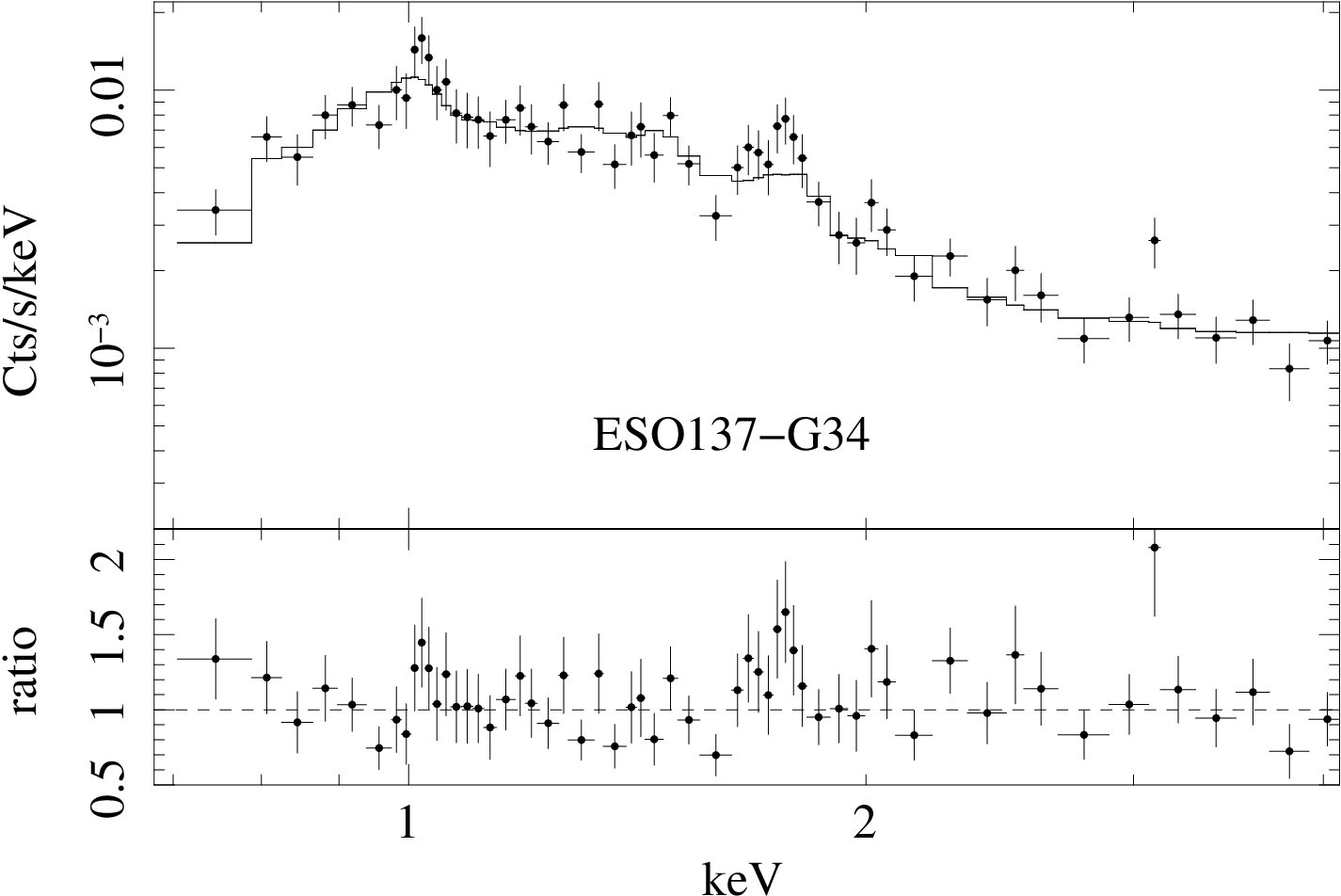}
\includegraphics[keepaspectratio=false, height=7.5cm, width=5.5cm,
angle=270]{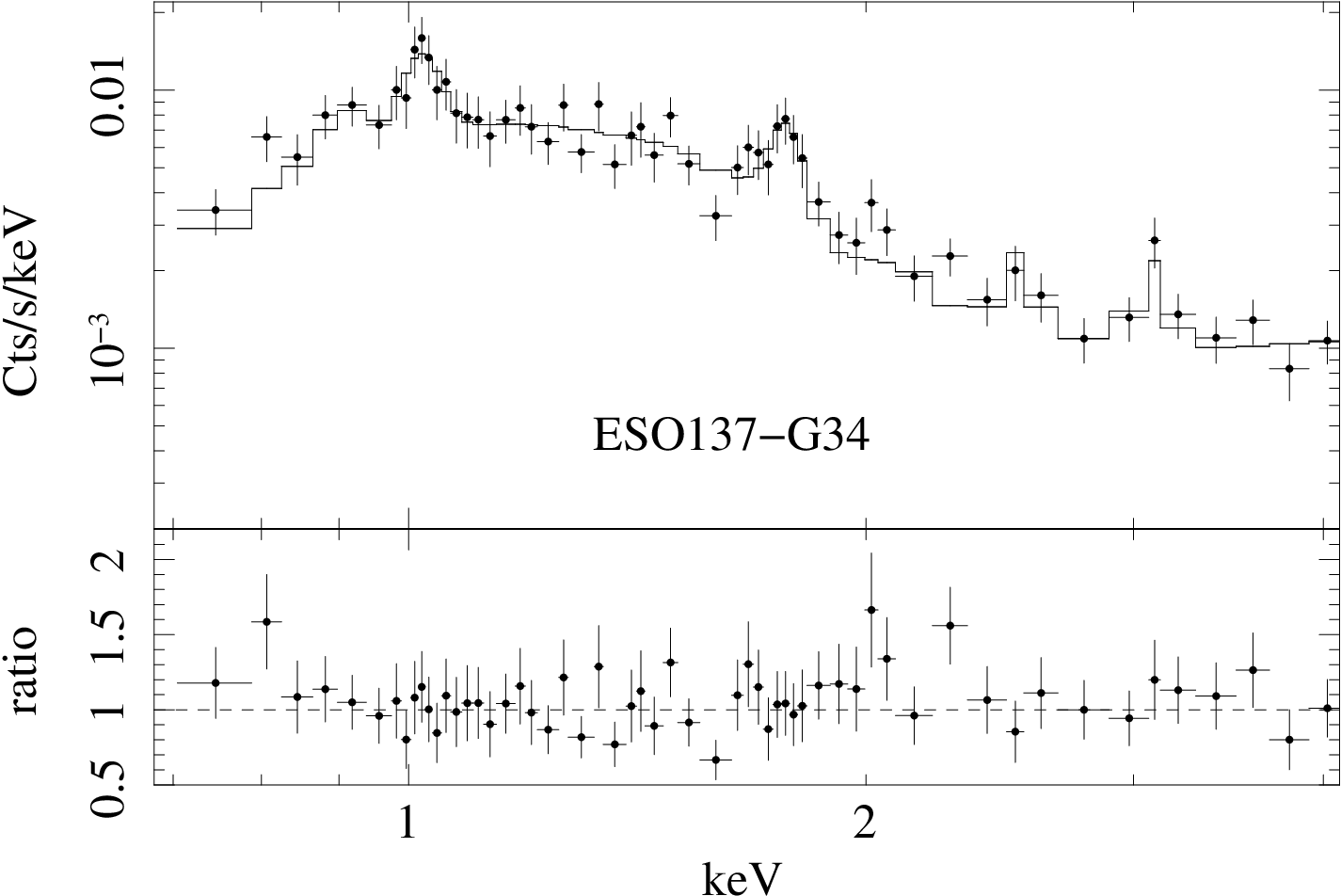}
\includegraphics[keepaspectratio=false, height=7.5cm, width=5.5cm,
angle=270]{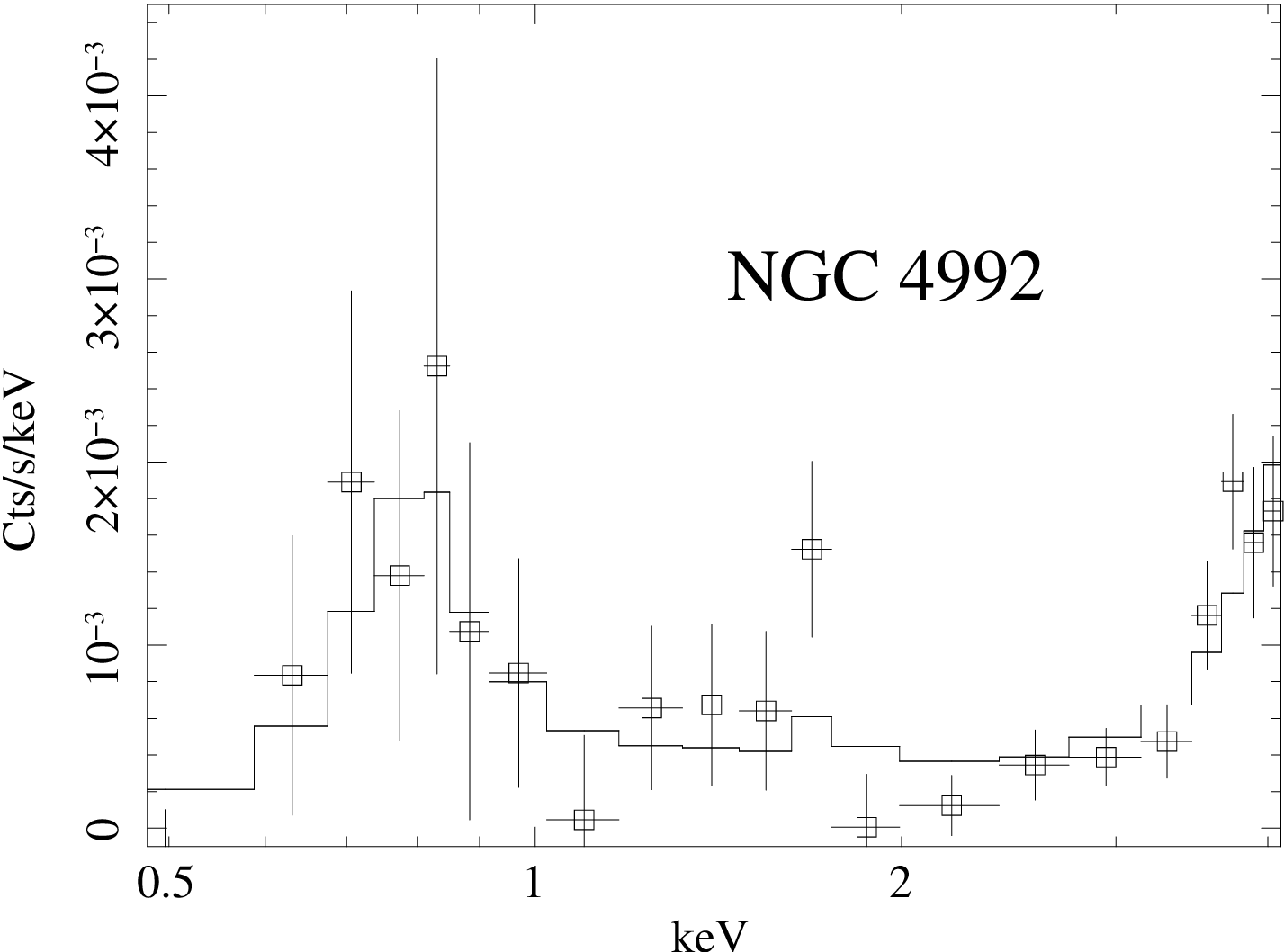}
\includegraphics[keepaspectratio=false, height=7.5cm, width=5.5cm,
angle=270]{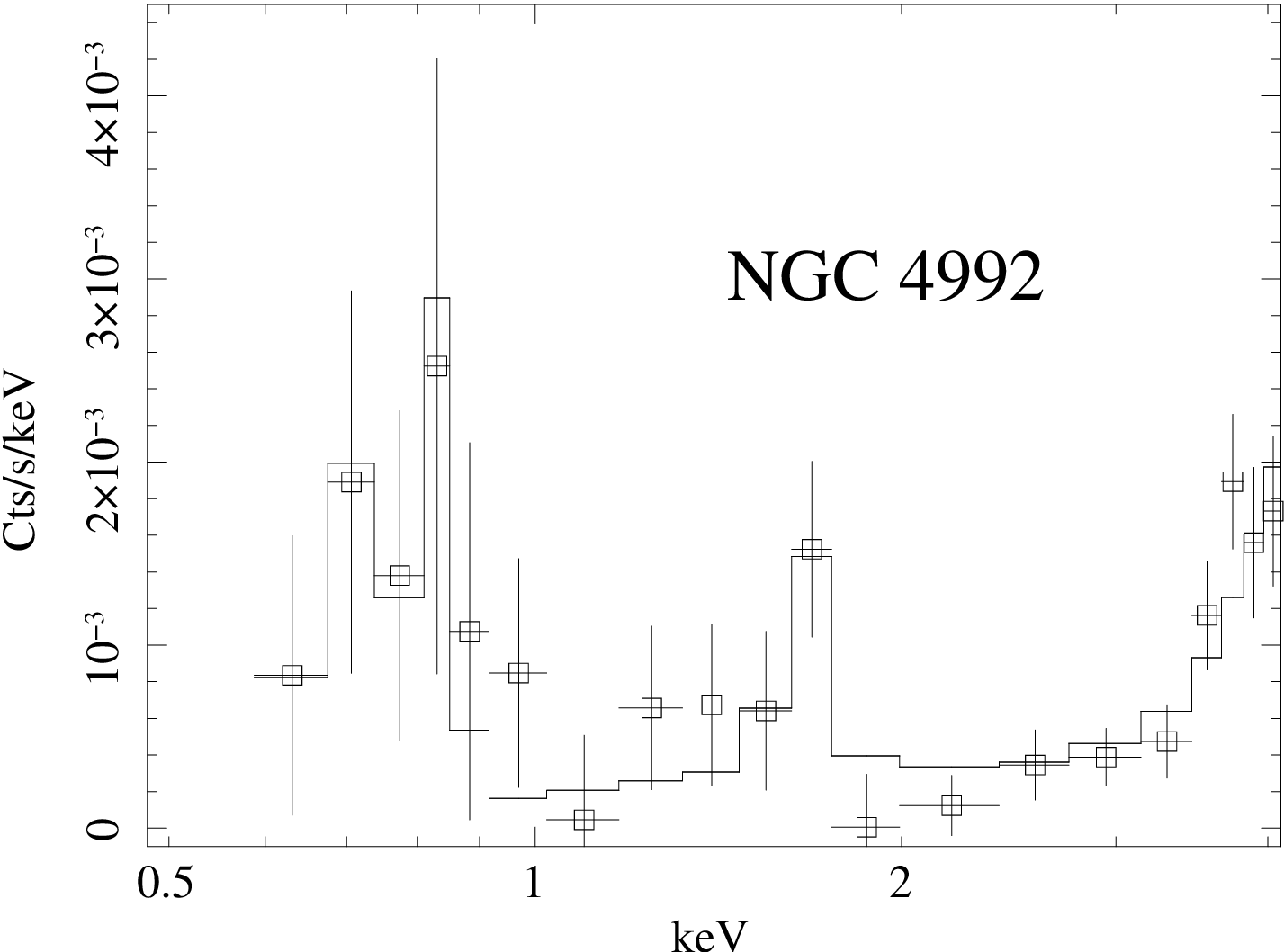}
\end{center}
\caption{The {\it Suzaku} soft X--ray spectra ($\lsimeq$ 4 keV) and data to model ratio 
of NGC 5728, ESO137--G34 and NGC 4992 (from top to bottom). In the left column 
the best fit to the soft X--ray spectrum obtained with a 
power law plus a thermal model is reported. In the right column the best fit with a power law
plus Gaussian lines is shown. For NGC 4992, only the soft X--ray spectrum, without residuals, is reported.} 
\label{oxspec}
\end{figure*}

\end{document}